\documentstyle[twoside,fleqn,epsfig,espcrc2]{article}

\newcommand{\beq}{\begin{equation}}
\newcommand{\eeq}{\end{equation}}
\newcommand{\Z}{{\sf Z \!\!\! Z}}

\title{Confinement without a center: the exceptional group $G(2)$}

\author{K.~Holland\address[UCSD]{Department of Physics, University of
    California at San Diego,
    La Jolla, CA 92093, U.S.A.},
  P.~Minkowski\address[Bern]{Institute for Theoretical Physics,
    Bern University, Sidlerstrasse 5, CH-3012 Bern, Switzerland},
  M.~Pepe\address[Paris]{Laboratoire de Physique Th\'eorique,
    Universit\'e de Paris-Sud,
    Bat. 210, F-91405 Orsay-Cedex, France}\thanks{Speaker at the
    Conference.},
  U.-J.~Wiese\addressmark[Bern]\thanks{On leave from MIT.}}

\begin{document}

\begin{abstract}
We discuss theories with the exceptional centerless gauge group $G(2)$,
paying attention to confinement and the pattern of
chiral symmetry breaking. Exploiting the Higgs mechanism to
break the symmetry down to $SU(3)$, we also present how the familiar
features of confinement and chiral symmetry breaking of $SU(3)$ gauge
theories reemerge. $G(2)$ gauge theories show up as an unusual
theoretical framework to study $SU(3)$ gauge theories without the
``luxury'' of a center. 
\end{abstract}

\maketitle
\section{Introduction and motivations}
\vspace{-.2cm}
In the last few years, accumulating numerical evidence of the relevance
of center vortices in the effective mechanism of confinement in
non--Abelian gauge theories has been collected. Center vortices and
twist sectors \cite{tHooft} are present in a pure gauge theory with
symmetry group $G$ if $\Pi_1 ( G/\mbox{center}(G) ) \neq {0}$, $\Pi_1$ being
the first homotopy group. The exceptional group $G(2)$ is the simplest
group without center vortices and twist sectors. Thus, it is
interesting to investigate how confinement can show up in a theory
with such a gauge group. Moreover, a property making $G(2)$ 
particularly interesting is that it has $SU(3)$ as a subgroup. 
In our study we have focused our attention on the way confinement and
the pattern of chiral symmetry breaking show up in $G(2)$ gauge theories and how
they change into the more familiar $SU(3)$ case as the symmetry gets
broken. 
\vspace{-.2cm}
\section{$G(2)$: basic generalities}
\vspace{-.2cm}
$G(2)$ is a subgroup of $SO(7)$. Its fundamental representation (rep) \{7\} is
7 dimensional and a matrix $\Omega$ satisfies the following
constraints:
\beq\label{SO7}
\det \Omega =1 \;\; ;\;\; 
\delta_{ab} = \delta_{a'b'}\; \Omega_{aa'}\; \Omega_{bb'}
\eeq
\beq\label{cubic}
\mbox{T}_{abc} = \mbox{T}_{a'b'c'}\; \Omega_{aa'}\; \Omega_{bb'}\;
\Omega_{cc'}
\;\; ;
\;\;\mbox{T}=\mbox{antisym}.\;\;
\eeq
In addition to the two defining $SO(7)$ properties (\ref{SO7}),  
$G(2)$ leaves invariant a completely
antisymmetric three-index tensor $\mbox{T}$ and is generated by 14 of the 21
$SO(7)$ generators. $G(2)$ has rank 2 and so its reps can
be drawn on a plane. For instance, this is the diagram of the fundamental one \{7\}:
\begin{figure}[htb]
\vspace{-1.2cm}
\begin{center}
\epsfig{file=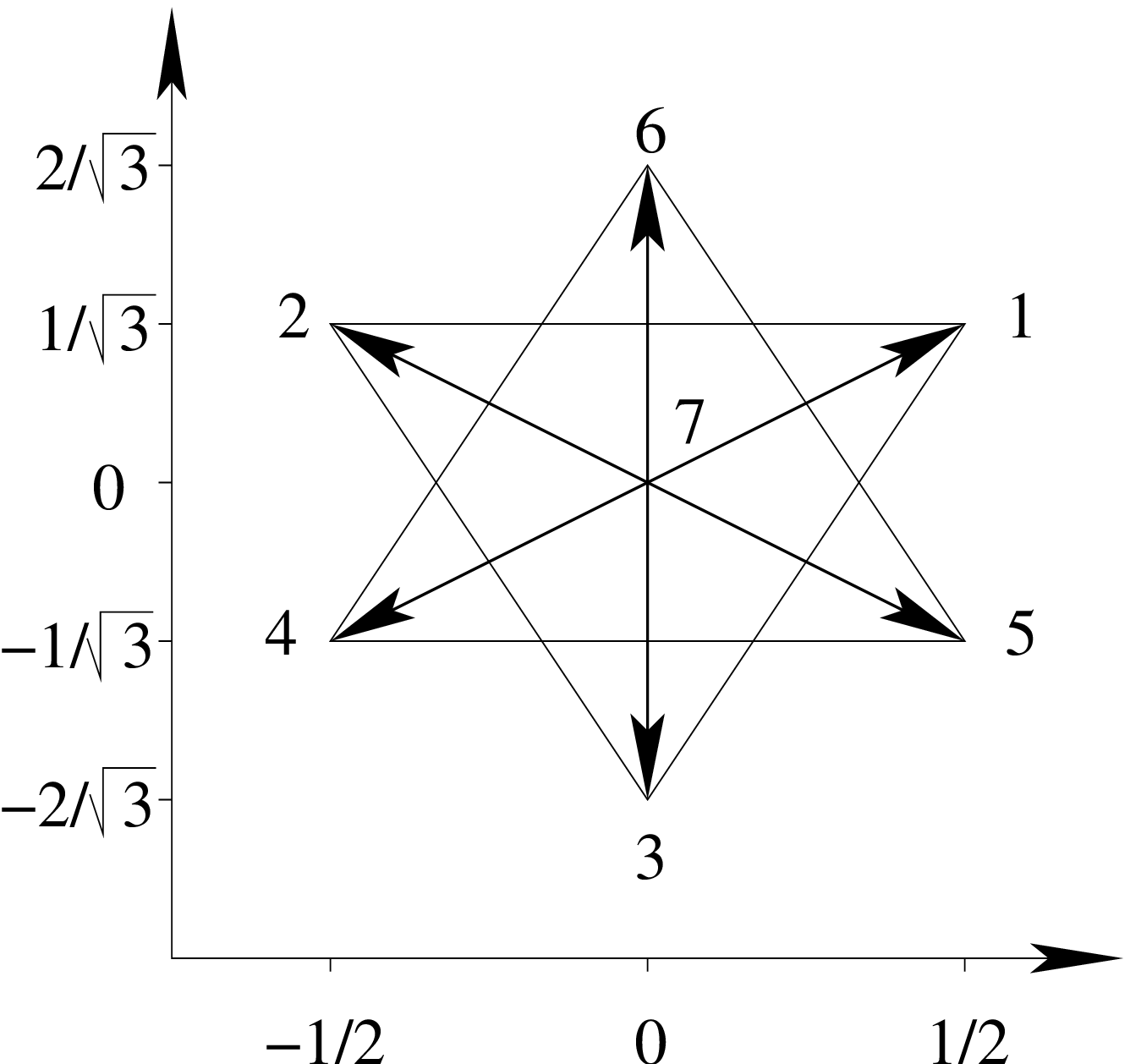,width=36mm,angle=0}
\vspace{-1.cm}
\caption{\it The weight diagram of the 7-dimensional fundamental representation
of $G(2)$.}
\end{center}    
\vspace{-1.1cm}
\end{figure}

$G(2)$ has $SU(3)$ as a subgroup in
the real reducible rep $\{3\} \oplus\{\overline 3\} \oplus \{1\}$. In
a suitable basis, 8 of the 14 $G(2)$ generators can be written in the
following way
\begin{equation} \label{su3gen}
\Lambda_a = \frac{1}{\sqrt{2}} \left( \begin{array}{ccc} \lambda_a & 0 & 0 \\ 
0 & \; -\lambda_a^* & 0\\ 0 & 0 & 0 \
\end{array} \right).
\end{equation}
where $\lambda_a$, $a=1,\ldots , 8$ are the usual $3 \times 3$
Gell--Mann matrices. The group $G(2)$ inherits from $SO(7)$ the
property of having real reps and so $G(2)$ ``quarks'' are equivalent
to $G(2)$ ``antiquarks''. $G(2)$ has a trivial center and there is no
feature similar to ``$N$--ality'', holding for $SU(N)$
groups. All $G(2)$ reps mix together and,
consequently, a heavy $G(2)$ ``quark'' can be screened by $G(2)$
``gluons''. Thus, the string breaks already in the pure glue sector without
the need for dynamical $G(2)$ ``quarks''. Finally, similarly to
$SU(3)$, in $G(2)$ Yang--Mills theory there are instantons --- $\Pi_3(G(2)) = \Z$
--- and two kinds of monopoles --- $\Pi_2(G(2)/U(1)^2) = \Z^2$.
\vspace{-.2cm}
\section{$G(2)$ Yang--Mills}
\vspace{-.2cm}
Let us first consider $G(2)$ Yang--Mills (YM) theory. There are 14
$G(2)$ ``gluons'' which transform according to the adjoint 
rep \{14\}. Restricting to the $SU(3)$ subgroup, this
rep is reducible and splits up into the sum
$\{8\} \oplus \{3\} \oplus\{\overline 3\}$. Thus, 8 of the 14 $G(2)$
``gluons'' are related among themselves as gluons and the remaining 6 
$G(2)$ ``gluons'' form a triplet and an antitriplet. These
last 6 $G(2)$ ``gluons'' behave like quarks and antiquarks w.r.t.~the
$SU(3)$ color degrees of freedom. The $G(2)$ YM theory is asymptotically free
and we expect it to be confining in the low energy regime but with a
vanishing string tension (defined as the slope of the heavy quark
potential when the distance goes to infinity), since the string breaks by pair production of
dynamical $G(2)$ ``gluons''. So, contrary to $SU(3)$
YM theory, in $G(2)$ YM theory we expect confinement
to resemble more closely that in QCD but without the complications related
to dynamical fermions. Due to the screening of $G(2)$ ``quarks'' by
$G(2)$ ``gluons'',  the Wilson loop is not a good order parameter for
confinement. At $T=0$, one can consider the Fredenhagen--Marcu
operator \cite{FM} as an order parameter for confinement. By strong
coupling computation, one obtains confining behaviour in this regime.
$SU(N)$ YM theory has a deconfinement phase transition at finite
temperature. Confined and deconfined phases differ by the way the center
symmetry is realized. In $G(2)$ YM, since the center 
is trivial, it is unclear how to define an order parameter to investigate
the issue of a finite temperature phase transition.
\vspace{-.2cm}
\section{$G(2)$ Yang--Mills + Higgs \{7\}}
\vspace{-.2cm}
Let us now break the $G(2)$ gauge symmetry to $SU(3)$. This can be
accomplished by adding to the $G(2)$ YM theory a Higgs field
in the fundamental rep \{7\} of $G(2)$. The 8 $G(2)$
``gluons'' related among themselves as gluons stay massless while the
remaining 6 get a mass proportional to the v.e.v.~of the Higgs
field. If the mass of these $G(2)$ ``gluons'' is not too high, they
participate in the dynamics but, as the v.e.v.~of the Higgs field
increases, they progressively decouple and, in the end, we are left with an $SU(3)$
gauge theory. Thus, a Higgs field in the rep \{7\}
gives us a handle to smoothly interpolate between
$G(2)$ and $SU(3)$. The breaking of the string between two heavy $G(2)$
``quarks'' happens for the pair production of these 6 massive
$G(2)$ ``gluons'' and so the breaking scale is related to their mass. 
As this mass increases, the breaking scale gets larger as well.
When the 6 massive $G(2)$ ``gluons'' completely decouple, it is sent
to infinity and we recover the familiar picture of the unbreakable
$SU(3)$ string.
\vspace{-.2cm}
\section{$G(2)$ QCD}
\vspace{-.2cm}
Let us now add to the $G(2)$ YM theory $N_f$ flavours of Majorana
fermions in the fundamental rep \{7\}. In this way, we
obtain a theory like QCD but with gauge group $G(2)$. As above, we
will exploit the Higgs mechanism to smoothly interpolate between
$G(2)$ and $SU(3)$. When we break the $G(2)$ symmetry to $SU(3)$, the
rep \{7\} reduces to the sum of  \{3\},
$\{\overline{\mbox{3}}\}$ and a color singlet. Thus, 
reexpressing the Majorana degrees of freedom in the following way 
\beq
\left(
\begin{array}{c}
\psi_M^{(1)}\\
\psi_M^{(2)}\\
\psi_M^{(3)}\\
\psi_M^{(4)}\\
\psi_M^{(5)}\\
\psi_M^{(6)}\\
\psi_M^{(7)}\\
\end{array}
\right)
--->
\left(
\begin{array}{c}
\psi_D^{(1)}=\psi_M^{(1)} + i \psi_M^{(4)}\\
\psi_D^{(2)}=\psi_M^{(2)} + i \psi_M^{(5)}\\
\psi_D^{(3)}=\psi_M^{(3)} + i \psi_M^{(6)}\\
^c\psi_D^{(1)}=\psi_M^{(1)} - i \psi_M^{(4)}\\
^c\psi_D^{(2)}=\psi_M^{(2)} - i \psi_M^{(5)}\\
^c\psi_D^{(3)}=\psi_M^{(3)} - i \psi_M^{(6)}\\
\psi_M^{(7)}=\psi_M^{(7)}\\
\end{array}
\right)
\eeq
we see that, by the Higgs mechanism, we are interpolating
between a $G(2)$ QCD--like theory with $N_f$ Majorana fermions and
QCD with $N_f$ Dirac quark flavours plus one Majorana fermion. However, this 
last particle is an $SU(3)$ color singlet and so does not feel the $SU(3)$
strong interactions. We now consider the issue of the pattern of
chiral symmetry breaking in $G(2)$ QCD and how this pattern interpolates to that of
QCD as the v.e.v.~of the Higgs field becomes large. Let us discuss
separately the cases $N_f =1$ and $N_f \geq 2$.\\
{$\bullet \bf {N_f =1}$}. 
Consider first the case of one flavour. In QCD the baryon number symmetry is 
$U(1)_{L=R}$. It stays unbroken and there is no Goldstone particle. In
$G(2)$ QCD, due to the reality of the $G(2)$ reps, $G(2)$
``quarks'' and $G(2)$ ``antiquarks'' are indistinguishable. Left ($L$) and
right ($R$) components do not transform independently but
$L=R^*$. So the baryonic $U(1)_{L=R}$ symmetry of QCD becomes 
$U(1)_{L=R=R^*}=\Z_B (2)$ in $G(2)$ QCD and the number of $G(2)$
``quarks'' is conserved only modulo two. Thus we have two kinds of states: those with an
odd (uGGG, uuu, ....) and those with an even (uu, ....) number of
$G(2)$ ``quarks'' u, bound or not with $G(2)$ ``gluons'' G. If we now add
the Higgs field to the dynamics, 6 $G(2)$ ``gluons'' become massive
and one can show that the states uGGG start to become heavy. The
mixing between $G(2)$ ``quarks'' and $G(2)$ ``antiquarks'' -- which is
mediated by the massive $G(2)$ ``gluons'' -- becomes weaker and baryon
number violating processes are rare. Then the $U(1)_{L=R} = U(1)_B$ symmetry 
of QCD reemerges as an approximate symmetry, becoming exact when the 6
massive $G(2)$ ``gluons'' decouple from the dynamics.\\
{$\bullet \bf{N_f \geq 2}$}. 
Let us now consider the case of two or more flavours. The Abelian part
of the chiral symmetry is that discussed in the $N_f =1$ case, so in
the following we will only take into account the non--Abelian part. In QCD the
pattern of chiral symmetry breaking is
$SU(N_f)_L \otimes SU(N_f)_R \rightarrow SU(N_f)_{L=R}$ with
$(N_f^2 -1)$ Goldstone bosons. Again, in $G(2)$ QCD, left and right
components are not independent and are related by $L=R^*$. So the
unbroken chiral symmetry is $SU(N_f)_{L=R^*}$ which breaks down to the
vector subgroup $SU(N_f)_{L=R=R^*} = SO(N_f)$. As a consequence, there
are $N_f (N_f + 1)/2 - 1$ Goldstone bosons, one for every broken
generator. As before, we add a Higgs field in the
fundamental rep \{7\} in order to smoothly
interpolate between $G(2)$ QCD and QCD. It is now better to start
from the QCD case -- i.e.~we assume a very large v.e.v.~for the
Higgs field -- and move to $G(2)$ QCD by decreasing this value.
$(N_f-1)$ of the $(N_f^2-1)$ Goldstone bosons of QCD are
self-conjugate (e.g. $\pi^0$ for $N_f=2$) and the remaining 
$2 [N_f (N_f - 1)/2]$ ones are
pairwise conjugate (e.g. $\pi^\pm$  for $N_f=2$). One can consider
linear combinations of these pairwise conjugated particles to build up states which
are even (e.g. $\pi^+ + \pi^-$) and odd (e.g. $\pi^+ - \pi^-$) under
charge conjugation. The odd states are invariant under $SO(N_f)$ and so
acquire a mass as the v.e.v.~of the Higgs field becomes smaller and
smaller, that is as the broken symmetry reduces from $SU(N_f)_{L=R}$
to $SU(N_f)_{L=R=R^*}$. The other 
$(N_f-1) + N_f (N_f - 1)/2 = N_f (N_f + 1)/2 - 1$ ones stay instead massless.
\vspace{-.2cm}
\section{Conclusions}
\vspace{-.2cm}
We have studied $G(2)$ gauge field theories with and without fermions. We have
focused our attention on confinement and the pattern of chiral
symmetry breaking. Adding a Higgs field in the fundamental
rep \{7\}, we can smoothly interpolate between $G(2)$ and
$SU(3)$. In particular, we have discussed how the familiar confinement
and pattern of chiral symmetry breaking of $SU(3)$ gauge theories
reemerge as the $G(2)$ gauge symmetry gets broken. In conclusion, we
have considered $G(2)$ theories as a theoretical laboratory to study
$SU(3)$ gauge theories in an unusual context and without the
``luxury'' of a center. In a forthcoming
paper, we will report on our study with more details. In that paper,
we will also discuss the 
$G(2)$ YM theory with 1 Majorana fermion flavour in the adjoint
rep \{14\} (SUSY--$G(2)$) and present analytic results in the
strong coupling approximation to support our heuristic
investigations.\\
{\it Acknowledgments.} M.P.~wishes to thank S.~Caracciolo, Ph.~de~Forcrand,
R.~Ferrari and O.~Jahn for useful discussions. M.P.~is supported by the
European Union Human Potential Program under  contract HPRN-CT-2000-00145,
Hadrons/Lattice QCD. K.H.~is supported by the US DOE under grant 
DOE-FG03-97ER40546.


\vspace{-.2cm}
\begin{thebibliography}{9}
\bibitem{tHooft}
G.~'t Hooft,
Nucl.\ Phys.\ B {\bf 138} (1978) 1.
\bibitem{FM}
K.~Fredenhagen and M.~Marcu,
Phys.\ Rev.\ Lett.\  {\bf 56} (1986) 223.
\end{thebibliography}
\end{document}